\documentclass[
10pt,
twocolumn,          
prl,                         
superscriptaddress
]{revtex4-2}

\usepackage{amsmath,amssymb,bm}
\usepackage{braket}
\usepackage{graphicx}
\usepackage[version=4]{mhchem}
\usepackage{mathrsfs}
\usepackage{xcolor}
\usepackage{soul}

\usepackage[
colorlinks=true,
linkcolor=blue,
citecolor=blue,
urlcolor=blue
]{hyperref}

\begin{document}


\title{Effective-action formulation of $\mathbf{k}$-odd bianisotropy and unidirectional polarization rotation in nonlocal temporal metamaterials}

\author{Neng Wang}
\author{Guo Ping Wang}%
 \email{gpwang@szu.edu.cn}
\affiliation{%
 State Key Laboratory of Radio Frequency Heterogeneous Integration, College of Physics and Optoelectronic Engineering, Shenzhen University, Shenzhen 518060, China
}%

\date{\today}

\begin{abstract}
We derive the effective action of temporal metamaterials up to first order in spatial derivatives through a multiscale analysis of the electromagnetic Lagrangian density. We show that the first-order spatial correction gives rise to an effective bianisotropic coupling tensor that is odd in the wavevector $\mathbf{k}$. Remarkably, chiral, pseudochiral, and omega-type bianisotropic responses emerge as distinct directional manifestations of a single nonlocal coupling tensor, even when the underlying electric and magnetic response tensors remain symmetric. Furthermore, we establish a framework for analyzing Lorentz reciprocity based on the Hessian of the effective action. We demonstrate that Lorentz reciprocity is preserved in the presence of omega-type coupling but is broken when chiral or pseudochiral coupling emerges. This nonreciprocity manifests as the unidirectional rotations of normalized Stokes vectors incurred by Floquet-engineered chirality and pseudochirality.
\end{abstract}

\maketitle


\section{Introduction}
Temporal metamaterials, in which electromagnetic properties are modulated in time, provide new opportunities for controlling wave propagation through frequency conversion \cite{lee2018linear,zhou2020broadband,moussa2023observation,apffel2022frequency}, parametric amplification \cite{wang2018photonic,lyubarov2022amplified,li2023stationary,park2025spontaneous}, Floquet engineering \cite{hayran2021spectral,rizza2022short,schindler2025floquet} and synthetic rotation \cite{liberal2026synthetic,nasari2026observation}. Unlike conventional spatial structuring, temporal modulation enables the electromagnetic response to be dynamically reconfigured, giving rise to phenomena such as photonic time crystals and magnetless nonreciprocity \cite{lustig2018topological,tong2025observation,wang2025expanding,allard2026broadband,sounas2017non,zang2019nonreciprocal,guo2019nonreciprocal}. To establish a macroscopic description of these systems, Pacheco-Peña and Engheta introduced the effective-medium concept for temporally layered structures \cite{pacheco2020effective}. Further developments in temporal effective-medium theory have extended homogenization to space–time-modulated structures \cite{huidobro2021homogenization}, dispersive media \cite{zhu2025temporal}, and higher-order effective constitutive responses \cite{rizza2022nonlocal,wang2024nonlocal,wang2024achieving,gong2025anomalous}, providing a broader framework for understanding the macroscopic behavior of dynamically modulated electromagnetic systems.

Beyond scalar modulation, temporally varying anisotropic media offer additional degrees of freedom for controlling polarization and electromagnetic coupling \cite{pacheco2020temporal,xu2021complete,rizza2023spin,mostafa2024temporal,mirmoosa2024time}. In particular, the temporal rotation of anisotropic material tensors provides a route to engineering Floquet-mediated interactions between electromagnetic modes. Recent studies have demonstrated diverse forms of dynamical polarization and spin control. Jaffray et al. reported nonreciprocal optical activity in photoexcited aluminum-doped zinc oxide \cite{jaffray2026all}, whereas Rizza et al. investigated spin conversion associated with phase conjugation at a space–time interface in a dispersive medium \cite{rizza2026optical}. In our recent work, we demonstrated a temporal Faraday effect arising from Floquet-induced chirality in a system with temporally modulated electric and magnetic principal axes \cite{wang2026temporal}. This finding raises a broader theoretical question: how can the interplay between temporal modulation and spatial dispersion be systematically described, and how are the resulting bianisotropic responses related to Lorentz reciprocity?

Here, we address these questions by deriving the effective electromagnetic action of temporally modulated anisotropic media through a multiscale expansion of the microscopic Lagrangian density. The first-order spatial correction gives rise to a nonlocal bianisotropic coupling tensor that is odd in the wavevector. We demonstrate that chiral, pseudochiral, and omega-type responses emerge as distinct directional manifestations of this unified tensor, even when the underlying electric and magnetic response tensors remain symmetric. Furthermore, we use the Hessian of the effective action to establish a systematic framework for examining Lorentz reciprocity. Our analysis distinguishes the reciprocity properties associated with different bianisotropic coupling configurations and connects the emergence of Floquet-induced chirality and pseudochirality to the unidirectional polarization rotations. These results provide a unified action-based framework for understanding nonlocal bianisotropy and reciprocity in temporal metamaterials.

\section{Multiscale analysis of the Lagrangian density}
Here, we consider that the temporal metamaterial are generally anisotropic with the permittivity and permeability tensors modulated periodically over time, \emph{i.e.}, $\hat{\varepsilon}(t)=\hat{\varepsilon}(t+T),\,\hat{\mu}(t)=\hat{\mu}(t+T)$, where $T$ is the modulation period. The constitutive tensors satisfy $\hat{\varepsilon}(t)=\hat{\varepsilon}^T(t),\,\hat{\mu}^T(t)=\hat{\mu}(t)$, namely the metamaterial is reciprocal at any instantaneous time. Then, the microscopic (unhomogenized) electromagnetic Lagrangian density reads as \cite{zhang2021electromagnetic}
\begin{align}\label{microLag}
    \mathcal{L}=\frac{1}{2}E_i\varepsilon_{ij}(t)E_j-\frac{1}{2}B_i\nu_{ij}(t)B_j,
\end{align}
where $\hat{\nu}=\hat{\mu}^{-1}$, and  $\mathbf{E},\,\mathbf{B}$ are the microscopic electromagnetic fields related to the vector potential $\mathbf{A}$ as
\begin{align}\label{gauge}
    E_i=-\partial_tA_i,\quad B_i=\epsilon_{ijk}\partial_jA_k,
\end{align}
with $\epsilon_{ijk}$ being the Levi-Civita symbol. Here, Cartesian Einstein notations are used and due to the source-free conditions, the gauge $\phi=0$ is applied for simplicity. Substituting Eqs. \eqref{microLag} and \eqref{gauge} into the action $\mathcal{S}[\mathbf{A}]=\int dr^3dt\mathcal{L}$, applying the stationary action principle $\delta\mathcal{S}[\mathbf{A}]=0$ and combining Eq. \eqref{gauge}, we obtain the familiar Maxwell's equations:
\begin{align}\label{maxwell}
    \partial_t[\hat{\varepsilon}(t)\mathbf{E}]-\nabla\times[\hat{\nu}(t)\mathbf{B}]=0,\quad
    \partial_t\mathbf{B}+\nabla\times\mathbf{E}=0.
\end{align}

To obtain the macroscopic (homogenized) Lagrangian density $\mathcal{L}_{\rm eff}$, we introduce the fast temporal coordinate $\tau=t/\eta$, where $\eta\ll 1$ is the dimensionless scale-separation parameter, and perform a multiscale analysis \cite{wang2024nonlocal,wang2024achieving,ciattoni2015nonlocal} about the microscopic Lagrangian density $\mathcal{L}$. Treating $\tau$ as an independent variable to $t$, the following operator transformations
\begin{align}\label{operator}
    \partial_t\to D_t=\partial_t+\eta^{-1}\partial_\tau,\quad \nabla\to\nabla
\end{align}
are adopted. 

We can expand the vector potential $\mathbf{A}$ in Maclaurin series of $\eta$ as
\begin{align}
    \mathbf{A}=\mathbf{A}_0(\mathbf{r},t)+\eta\mathbf{A}_1(\mathbf{r},t,\tau)+\eta^2\mathbf{A}_2(\mathbf{r},t,\tau)+\dots,
\end{align}
where the gauge normalization 
\begin{align}\label{gaugenorm}
    \mathbf{A}_0=\langle\mathbf{A}\rangle, \quad \langle \mathbf{A}_{n\ge1}\rangle=0,
\end{align}
are used, with the temporal average henceforth defined as
\begin{align}
    \langle f\rangle =\frac{1}{\mathcal{T}}\int_0^\mathcal{T}f(t,\tau)d\tau,\quad \mathcal{T}=T/\eta.
\end{align}
Eq.~\eqref{gaugenorm} implies that the temporal averages of all orders are grouped into the zeroth order $\mathbf{A}_0$. Combining Eqs.~\eqref{gauge} and \eqref{operator}, 
\begin{align}\label{AtoEB}
    \mathbf{E}=-D_t\mathbf{A}=\sum_{j=0}^\infty \eta^j\mathbf{E}_j,\quad \mathbf{B}=\nabla\times\mathbf{A}=\sum_{j=0}^\infty \eta^j\mathbf{B}_j,
\end{align}
where 
\begin{align}\label{maclaurinE}
    \mathbf{E}_j=-\partial_t\mathbf{A}_j-\partial_\tau\mathbf{A}_{j+1},\quad \mathbf{B}_j=\nabla\times\mathbf{A}_j.
\end{align}
From Eq.~\eqref{AtoEB}, we acknowledge that the zeroth order $\mathbf{A}_0$ must be $\tau$-independent, otherwise the lowest order of $-D_t\mathbf{A}$, namely $-\eta^{-1}\partial_\tau\mathbf{A}_0$, will diverge as $\eta\to0$. Using Eq.~\eqref{gaugenorm}, the macroscopic electromagnetic fields are easily obtained as
\begin{subequations}\label{macroEB}
\begin{equation}
    \overline{\mathbf{\mathbf{E}}}=-\langle D_t\mathbf{A}\rangle=-\partial_t\mathbf{A}_0,
    \end{equation}    
    \begin{equation}
    \overline{\mathbf{B}}=\langle\nabla\times\mathbf{A}\rangle=\nabla\times\mathbf{A}_0=\mathbf{B}_0.
    \end{equation}
\end{subequations}

Inserting \eqref{maclaurinE} and \eqref{macroEB} into $\mathcal{L}$, we obtain 
\begin{align}
    \mathcal{L}=\mathcal{L}_0+\eta\mathcal{L}_1+o(\eta^2),
\end{align}
where 
\begin{subequations}
\begin{equation}
    \mathcal{L}_0=\frac{1}{2}\mathbf{E}_0\cdot\hat{\varepsilon}(\tau)\mathbf{E}_0-\frac{1}{2}\mathbf{B}_0\cdot\hat{\nu}(\tau)\mathbf{B}_0,
    \end{equation}
    \begin{equation}
      \mathcal{L}_1=\mathbf{E}_0\cdot\hat{\varepsilon}(\tau)\mathbf{E}_1-\mathbf{B}_0\cdot\hat{\nu}(\tau)\mathbf{B}_1.  
    \end{equation}    
\end{subequations}
Substituting Eq.~\eqref{operator} into the first of Eq.~\eqref{maxwell}, the leading order is given by
\begin{align}
    \partial_\tau [\hat{\varepsilon}(\tau)(\partial_t\mathbf{A}_0-\partial_\tau\mathbf{A}_1)]=0.
\end{align}
Thus, $\hat{\varepsilon}(\tau)(\partial_t\mathbf{A}_0-\partial_\tau\mathbf{A}_1)$ is $\tau$-independent, leading to
\begin{align}\label{zero}
    \langle\hat{\varepsilon}(\tau)(\partial_t\mathbf{A}_0-\partial_\tau\mathbf{A}_1)\rangle=-\hat{\varepsilon}(\tau)\mathbf{E}_0=-\overline{\mathbf{D}}_0(t). 
\end{align}
Therefore, the temporal average of $\mathcal{L}_0$ is calculated as
\begin{align}
    \langle\mathcal{L}_0\rangle=&\frac{1}{2}\langle\mathbf{E}_0\rangle\cdot \overline{\mathbf{D}}_0-\frac{1}{2}\overline{\mathbf{B}}\cdot\langle\hat{\nu}(\tau)\rangle\overline{\mathbf{B}}\nonumber\\=&\frac{1}{2}\overline{\mathbf{E}}\cdot\hat{\varepsilon}_{e}\overline{\mathbf{E}}-\frac{1}{2}\overline{\mathbf{B}}\cdot\hat{\nu}_{e}\overline{\mathbf{B}},
\end{align}
where we have denoted
\begin{align}\label{effecepsmu}
    \hat{\varepsilon}_e=(\langle\hat{\varepsilon}^{-1}(\tau)\rangle)^{-1},\quad \hat{\nu}_e=\langle\hat{\nu}(\tau)\rangle.
\end{align}

The combination of Eqs.~\eqref{macroEB} and \eqref{zero} yields
\begin{align}
    \partial_\tau\mathbf{A}_1=(\hat{\varepsilon}^{-1}(\tau)\hat{\varepsilon}_{\rm e}-\hat{I}_3)\overline{\mathbf{E}},
\end{align}
where $\hat{I}_3$ is the $3\times3$ identity matrix. Introducing the temporal corrector $\hat{Q}(\tau)$, which satisfies
\begin{align}\label{Q}
    \partial_\tau\hat{Q}(\tau)=\hat{\varepsilon}^{-1}(\tau)\hat{\varepsilon}_{\rm e}-\hat{I}_3,
\end{align}
and 
\begin{align}\label{Q1}
\langle \hat{Q}(\tau)\rangle=0,
\end{align}
we obtain $\mathbf{A}_1=\hat{Q}(\tau)\overline{\mathbf{E}}$ and 
\begin{align}\label{B1}
    \mathbf{B}_1=\nabla\times\mathbf{A}_1=\nabla\times(\hat{Q}(\tau)\overline{\mathbf{E}}).
\end{align}
The constraint in Eq.~\eqref{Q1} is due to $\langle\mathbf{A}_1\rangle=\langle \hat{Q}(\tau)\rangle \overline{\mathbf{E}}=0$, as per Eq.~\eqref{gaugenorm}. Inserting Eq.~\eqref{B1} into $\mathcal{L}_1$, and reminding that $\hat{\varepsilon}^T(\tau)=\hat{\varepsilon}(\tau)$ and $\langle\mathbf{E}_1\rangle=0$, the temporal average of $\mathcal{L}_1$ is then given by
\begin{align}
    \langle\mathcal{L}_1\rangle=&\langle\mathbf{E}_0\cdot\hat{\varepsilon}(\tau)\mathbf{E}_1\rangle-\langle\mathbf{B}_0\cdot\hat{\nu}\mathbf{B}_1\rangle\nonumber\\=&\langle\hat{\varepsilon}^T(\tau)\mathbf{E}_0\cdot\mathbf{E}_1\rangle-\langle\mathbf{B}_0\cdot\hat{\nu}(\tau)\mathbf{B}_1\rangle\nonumber\\=&\overline{\mathbf{D}}_0\cdot\langle\mathbf{E}_1\rangle-\overline{\mathbf{B}}\cdot\langle\hat{\nu}(\tau)\nabla\times\hat{Q}(\tau)\overline{\mathbf{E}}\rangle\\=&-\overline{\mathbf{B}}\cdot\langle\hat{\nu}(\tau)\nabla\times\hat{Q}(\tau)\overline{\mathbf{E}}\rangle.\nonumber
\end{align}
Using the Cartesian Einstein notation and denoting 
\begin{align}
    C_{ikl}=\langle\nu_{ij}\epsilon_{jkm}Q_{ml}\rangle,
\end{align}
the effective Lagrangian density up to the first order of $\eta$ is expressed as
\begin{align}\label{effecLag}
    \mathcal{L}_{\rm eff}=&\langle\mathcal{L}\rangle=\langle\mathcal{L}_0\rangle+\langle\mathcal{L}_1\rangle\nonumber\\=&\frac{1}{2}\overline{E}_{i}\varepsilon_{{\rm e},ij}\overline{E}_j-\frac{1}{2}\overline{B}_i\nu_{{\rm e},ij}\overline{B}_j-\eta\overline{B}_iC_{ikl}\partial_k\overline{E}_l.
\end{align}
Eq.~\eqref{effecLag} is the key result of this work, based on which we will establish the effective medium framework and analyze the Lorentz reciprocity of the temporal metamaterial.

\section{Effective k-odd bianisotropy}
From \eqref{effecLag}, the effective constitutive relations of this temporal metamaterial can be obtained directly,
\begin{subequations}\label{effecconst}
  \begin{equation}
    \overline{D}_i=\frac{\delta \mathcal{L}_{\rm eff}}{\delta \overline{E}_i}=\frac{\partial \mathcal{L}_{\rm eff}}{\partial \overline{E}_i}-\partial_k\frac{\partial \mathcal{L}_{\rm eff}}{\partial (\partial_k \overline{E}_i)}=\varepsilon_{{\rm e}, ij}\overline{E}_j+\eta C_{kli}\partial_l \overline{B}_k,
    \end{equation}
    \begin{equation}
    \overline{H}_i=-\frac{\delta \mathcal{L}_{\rm eff}}{\delta \overline{B}_i}=-\frac{\partial \mathcal{L}_{\rm eff}}{\partial \overline{B}_i}=\nu_{{\rm e},{ij}}\overline{B}_j+\eta C_{ikl}\partial_k\overline{E}_l.
\end{equation}  
\end{subequations}
According to Eq.~\eqref{effecconst}, the first order nonlocal correction of $\mathcal{L}_{\rm eff}$ just contributes an effective bianisotropic coupling. Unlike the local scenario, this nonlocal bianisotropic coupling relates to the spatial gradients of effective fields rather than the effective fields themselves.

Specifically, for Fourier representation, where $\mathbf{E}(\mathbf{r}),\mathbf{B}(\mathbf{r})\propto e^{i\mathbf{k}\cdot\mathbf{r}}$, we have $\partial_j\to ik_j$, and the effective constitutive relations reduce to
\begin{align}
    \left(\begin{array}{c}\overline{\mathbf{D}}\\ \overline{\mathbf{H}}\end{array}\right)=\left(\begin{array}{cc}\hat{\varepsilon}_e & -i\eta\hat{C}_k^T\\ -i\eta\hat{C}_k & \hat{\nu}_e\end{array}\right)\left(\begin{array}{c}\overline{\mathbf{E}}\\ \overline{\mathbf{B}}\end{array}\right),
\end{align}
where 
\begin{align}
    (C_k)_{ik}=C_{ijk}k_j.
\end{align}
Therefore, the bianisotropic coupling tensor $\hat{C}_k$ is not only $\mathbf{k}$-dependent but also odd parity with $\mathbf{k}$.

\subsection{Necessary conditions for nonzero bianisotropic coupling}
If $\hat{\varepsilon}$ is static, \emph{i.e.}, $\hat{\varepsilon}(t)\equiv\hat{\varepsilon}_e$, the temporal corrector $\hat{Q}(\tau)$ vanishes directly as per Eq.~\eqref{Q}, making $C_{ikl}=\langle\nu_{ij}\epsilon_{jkm}Q_{ml}\rangle\to0$. And if $\hat{\nu}$ is static, using Eq.~\eqref{Q1},
\begin{align}
    C_{ikl}=\langle v_{ij}\epsilon_{jkm} Q_{ml}\rangle=v_{ij}\epsilon_{jkm}\langle Q_{ml}\rangle=0.
\end{align}
On the other hand, if there is a temporal origin $\tau_0$ such that 
\begin{align}
    \hat{\varepsilon}(\tau_0+\tau)=\hat{\varepsilon}(\tau_0-\tau),\quad \hat{\nu}(\tau_0+\tau)=\hat{\nu}(\tau_0-\tau),
\end{align}
namely the entire temporal modulation protocol obeys time-reversal symmetry about a certain temporal origin, then $\hat{Q}(\tau)$ is odd while $\hat{\nu}(\tau)$ is even about $\tau_0$, rendering 
\begin{align}\label{Cikl}
    C_{ikl}=\langle v_{ij}\epsilon_{jkm} Q_{ml}\rangle=0.
\end{align}
Therefore, an effective nonzero bianisotropic coupling requires simultaneous modulation of both the permittivity and permeability as well as breaking time-reversal symmetry of the system. However, these conditions are albeit necessary but not sufficient for realizing nonzero bianisotropic coupling. For example, if we let both $\hat{\varepsilon}(\tau)$ and $\hat{\nu}(\tau)$ to be modulated in sawtoothwave forms, then within a period, $\nu_{ij}$ is odd while $Q_{ml}$ is even about a certain temporal origin $\tau_0$, with time-reversal symmetry broken, $C_{ikl}$ is still vanishing.

\subsection{Direction-dependent bianisotropic character}
$\hat{C}_k$ is generally a $3\times3$ matrix. The irreducible decomposition of $\hat{C}_k$ yields,
\begin{align}\label{decompose}
    \hat{C}_k=&\frac{{\rm Tr}(\hat{C}_k)}{3}\hat{I}_3+\left[\frac{\hat{C}_k+\hat{C}_k^T}{2}-\frac{{\rm Tr}(\hat{C}_k)}{3}\hat{I}_3\right]\nonumber\\ &+\frac{\hat{C}_k-\hat{C}_k^T}{2},
\end{align}
where $\rm Tr(\cdot)$ denotes the trace. The first term on the right side of Eq.~\eqref{decompose} is a scalar matrix, denoting the isotropic chiral coupling. The second term in the square bracket is traceless and symmetric, representing the pseudochiral coupling. And the third term is traceless and antisymmetric, denoting the omega-type coupling \cite{mackay2010electromagnetic}. 

\begin{figure*}[t]
    \centering
    \includegraphics[width=0.8\linewidth]{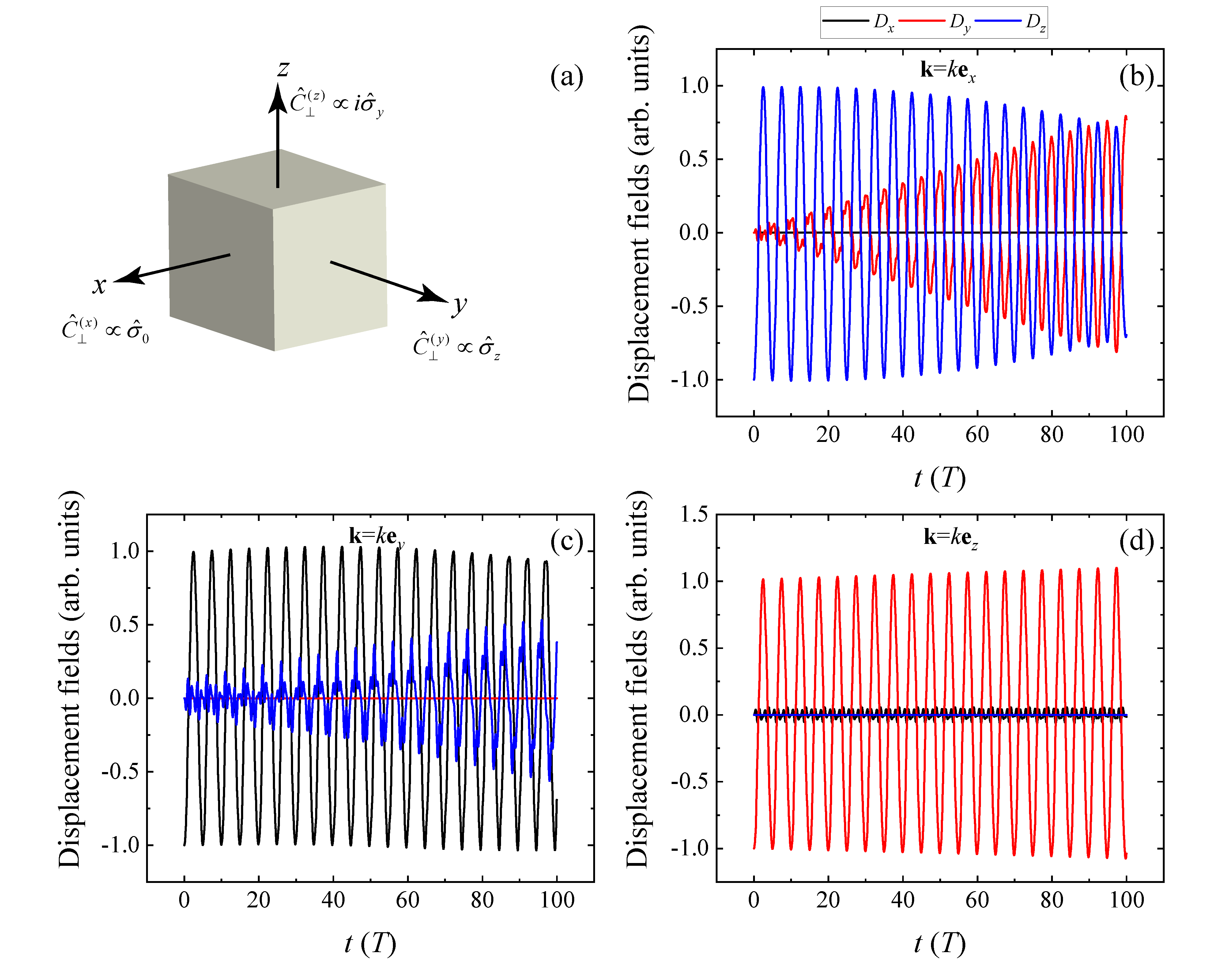}\caption{Direction-dependent bianisotropy. (a) Schematic of the direction-dependent bianisotropic character. (b)-(d) Temporal evolutions of the microscopic displacement fields ($D_x$ black, $D_y$ red and $D_z$ blue) of plane waves with (b) $\mathbf{k}=k\mathbf{e}_x$, (c) $\mathbf{k}=k\mathbf{e}_y$ and (d) $\mathbf{k}=k\mathbf{e}_z$, where $k=0.2\Omega/c$. The plane waves are initially linearly polarized. The parameters of the temporal metamaterial are given by Eq.~\eqref{sinu} with $\varepsilon_s=\varepsilon_0,\,\nu_s=\mu_0^{-1}$ and $\gamma=0.25$. }\label{fig1}
\end{figure*}

However, noting that for plane wave modes, the electromagnetic fields are polarized on the plane perpendicular to the wavevector $\mathbf{k}$, the effective constitutive tensor is then the $2\times2$ block on this transverse subspace. For instance, when $\mathbf{k}=k_x\hat{e}_x$, the effective part of $\hat{C}_k$ is the block on the transverse $yz$ plane, $\hat{C}_\perp^{(x)}$. Owing to that, we can well optimize the temporal profiles of the permittivity $\hat{\varepsilon}(\tau)$ and permeability $\hat{\mu}(\tau)$ tensors such that different bianisotropic couplings are along the three axes, \emph{e.g.}, isotropic chiral along $k_x$, pseudochiral along $k_y$ and omega-type along $k_z$, thus achieving direction-dependent bianisotropic character. This manifests mathematically as
\begin{align}\label{diredepen}
\hat{C}_\perp^{(x)}\propto k_x\hat{\sigma}_0,\quad & \hat{C}_\perp^{(y)}\propto k_y(a_1\hat{\sigma}_x+a_2\hat{\sigma}_z), \nonumber \\ &\hat{C}_\perp^{(z)}\propto ik_z\hat{\sigma}_y,
\end{align}
where $\hat{\sigma}_0$ is the $2\times2$ identity matrix, $\hat{\sigma}_{x,y,z}$ are the Pauli matrices and $a_1,\,a_2$ are two arbitrary constants.

As a typical example, we consider the following sinusoidal modulation
\begin{subequations}\label{sinu}
    \begin{equation}
        \hat{\nu}(\tau)=\nu_s+\gamma\mu_0^{-1}\hat{M}_1\sin(\eta\Omega \tau),
    \end{equation}
    \begin{equation}
       \hat{\varepsilon}^{-1}(\tau)=\varepsilon_s^{-1}[1+\gamma\hat{M}_2\cos(\eta\Omega \tau)], 
    \end{equation}
    \end{subequations}
where $\varepsilon_s,\,\nu_s$ are the static components, $\Omega=2\pi/T$ is the modulation frequency, $\gamma$ denotes the modulation strength and 
\begin{subequations}\label{M1M2}
    \begin{equation}
    \hat{M}_1=\left(\begin{array}{ccc}-1 & -1 & 1+\sqrt{2} \\ -1 & 1 & -1\\ 1+\sqrt{2} & -1 & -1\end{array}\right),
\end{equation}
\begin{equation}
    \hat{M}_2=\frac{1}{2}\left(\begin{array}{ccc}1 & 1 & 1-\sqrt{2} \\ 1 & -1 & -1\\ 1-\sqrt{2} & -1& 1\end{array}\right).
\end{equation}
\end{subequations}
As per Eq.~\eqref{sinu}, both the permittivity and permeability are modulated and there is a no temporal origin $\tau_0$ such that $\hat{\varepsilon}(\tau_0+\tau)=\hat{\varepsilon}(\tau_0-\tau),\,\hat{\nu}(\tau_0+\tau)=\hat{\nu}(\tau_0-\tau)$ are simultaneously fulfilled. And since $\hat{M}_1=\hat{M}_1^T,\,\hat{M}_2=\hat{M}_2^T$, $\hat{\varepsilon}(\tau)$ and $\hat{\nu}(\tau)$ are symmetric for all times. 

Substituting Eqs.~\eqref{sinu} and \eqref{M1M2} into Eqs.~\eqref{effecepsmu} and \eqref{Q}, we easily obtain that $\hat{\varepsilon}_e=\varepsilon_s\hat{I}_3,\,\hat{\nu}_e=\nu_s\hat{I}_3$, and $\hat{Q}(\tau)=(\eta\Omega)^{-1}\gamma\hat{M}_2\sin(\eta\Omega\tau)$. And using Eq.~\eqref{Cikl}, we obtain
\begin{align}\label{diredepen1}
    i\eta\hat{C}_\perp^{(x)}=&\frac{ik_x\gamma^2}{2\Omega\mu_0}\hat{\sigma}_0,\quad i\eta\hat{C}_\perp^{(y)}=\frac{ik_y\gamma^2}{2\Omega\mu_0}\hat{\sigma}_z,\nonumber \\ &i\eta\hat{C}_\perp^{(z)}=-\frac{k_z\gamma^2}{2\Omega\mu_0}\hat{\sigma}_y.
\end{align}
Thus, the bianisotropic couplings of isotropic chiral along $x$, pseudochiral along $y$ and omega-type along $z$ are achieved, with equal coupling magnitudes for the same wavenumber. 

Fig.~\ref{fig1} shows the temporal evolution of the microscopic electric displacement fields for linearly polarized waves propagating along the three principal axes. The microscopic fields are calculated numerically by iteratively solving Maxwell’s equations in Eq.~\eqref{maxwell} in the time domain. As illustrated in Fig.~\ref{fig1}(a), the temporal metamaterial exhibits direction-dependent bianisotropic responses. For propagation along the $x$-axis, the electric displacement field is initially polarized along $z$, while $D_y$ gradually increases, indicating continuous polarization conversion. For propagation along the $y$-axis, the linear polarization state progressively evolves because the principal optical axes are no longer aligned with the $x$- and $y$-axes. A detailed analysis of the polarization dynamics is presented in the next section. In contrast, for propagation along the $z$-axis, the plane wave largely preserves its linear polarization along $\mathbf{e}_y$, with $D_x$ and $D_z$ remaining negligibly small. This behavior is consistent with the linearly polarized eigenmodes supported by the omega-type effective medium.

These results establish that chiral, pseudochiral, and omega-type bianisotropic responses can emerge as different directional manifestations of a single nonlocal coupling tensor, even when the underlying electric and magnetic response tensors remain symmetric.

\section{Action Hessian}
Rewriting \(\mathbf{A}_0\) as \(\overline{\mathbf{A}}\) in Eq.~\eqref{macroEB}, the effective Lagrangian density in terms of the effective vector potential becomes
\begin{align}\label{LeffA}
    \mathcal{L}_{\rm eff}=&\frac{1}{2} (\partial_t\overline{A}_{i})\varepsilon_{{\rm e},ij}(\partial_t\overline{A}_{j})-\frac{1}{2}\epsilon_{imn}(\partial_m\overline{A}_{n})\nu_{{\rm e},ij}\epsilon_{jpq}(\partial_p\overline{A}_{q})\nonumber\\&+\eta\epsilon_{imn}C_{ijk}(\partial_m\overline{A}_{n})(\partial_i\partial_t\overline{A}_{k}).
\end{align}
Substituting Eq.~\eqref{LeffA} into the effective action $\mathcal{S}_{\rm eff}=\int dr^3dt \mathcal{L}_{\rm eff}$, and taking the second functional variation with respect to the effective vector potential yields the Hessian of the effective action \cite{toms2012schwinger}
\begin{align}
    \mathcal{H}_{ab}=&\frac{\delta^2\mathcal{S}_{\rm eff}}{\delta \overline{A}_a\delta\overline{A}_b}=-\varepsilon_{e,ab}\partial_t^2-\epsilon_{ima}\nu_{e,ij}\epsilon_{jpb}\partial_m\partial_p\nonumber\\&+\frac{\eta}{2}(\epsilon_{ima}C_{ikb}+C_{ika}\epsilon_{imb})\partial_m\partial_k\partial_t.
\end{align}

\subsection{Dispersion}
By the principle of stationary action, \(\delta\mathcal{S}_{\rm eff}=0\), the effective vector potential satisfies the Euler–Lagrange equation governed by the action Hessian: 
\begin{align}\label{linear}
   \mathcal{H}_{ab}\overline{A}_b=0. 
\end{align}
Applying the Fourier substitutions
\begin{align}
    \partial_t=-i\omega,\quad \partial_j=ik_j,
\end{align}
Eq.~\eqref{linear} can be expressed in Fourier space as
\begin{align}\label{linear1}
   H_{ab}(\omega,\mathbf{k})\overline{A}_b(\mathbf{k})=0, 
\end{align}
where $H_{ab}(\omega,\mathbf{k})$ denotes the Fourier-space action Hessian. In matrix form, it reads
\begin{align}\label{linear2}
   \hat{H}(\omega,\mathbf{k})=\omega^2\hat{\varepsilon}_e-\hat{K}^T\hat{\nu}_e\hat{K}+\frac{i\eta\omega}{2}(\hat{K}^T\hat{C}_k-\hat{C}_k^T\hat{K}),
\end{align}
with 
\begin{align}
    K_{ij}=\epsilon_{imj}k_m
\end{align}
defining the curl matrix. For a nontrivial solution \(\overline{\mathbf A}(\mathbf k)\) to exist, the Fourier-space Hessian must be singular. This yields the secular equation
\begin{align}\label{secu1}
    \det\hat{H}(\omega,\mathbf{k})=0,
\end{align}
whose roots determine the dispersion relations \(\omega(\mathbf{k})\) of the effective medium.

However, because Eq.~\eqref{secu1} is formulated in terms of the effective vector potential, the full Hessian includes two additional zero-frequency branches associated with gauge redundancy. To eliminate these unphysical degrees of freedom, we impose the Coulomb gauge, \(\nabla\cdot\overline{\mathbf{A}}=i\mathbf{k}\cdot\overline{\mathbf{A}}=0\), which restricts \(\overline{\mathbf{A}}\) to the two-dimensional transverse subspace for \(\mathbf{k}\neq 0\). We introduce two orthonormal transverse unit vectors, \(\mathbf{e}_1\) and \(\mathbf{e}_2\), satisfying \(\mathbf{e}_a\cdot\mathbf{k}=0\) and \(\mathbf{e}_a\cdot\mathbf{e}_b=\delta_{ab}\), and define the \(3\times2\) matrix \(\hat{U}=(\mathbf{e}_1,\mathbf{e}_2)\). The Hessian projected onto the transverse subspace is then
\begin{align}
\hat{H}_{\rm red}(\omega,\mathbf{k})
=\hat{U}^\dagger\hat{H}(\omega,\mathbf{k})\hat{U},
\end{align}
and the physical dispersion relations follow from 
\begin{align}
    \det\left|\hat{H}_{\rm red}(\omega,\mathbf{k})\right|=0.
\end{align}

Since both \(\hat{K}\) and \(\hat{C}_{\mathbf{k}}\) are odd functions of \(\mathbf{k}\), their products in Eq.~\eqref{linear2} are even in \(\mathbf{k}\). Consequently, the Fourier-space Hessian satisfies $\hat{H}(\omega,\mathbf{k})=\hat{H}(\omega,-\mathbf{k})$. Furthermore, for real-valued effective tensors, the Hessian obeys $\hat{H}(\omega,\mathbf{k})=\hat{H}^\dag(-\omega,\mathbf{k})$. These symmetries imply that the secular equation is invariant under both \(\mathbf{k}\rightarrow-\mathbf{k}\) and \(\omega\rightarrow-\omega\). Thus, the resulting dispersion spectrum is symmetric about both the \(\mathbf{k}=0\) and \(\omega=0\) axes.

\subsection{Lorentz reciprocity}
For a time-invariant, spatially homogeneous and spatially dispersive medium, Lorentz reciprocity is manifested in the symmetry of the Green's function \cite{xie2009general}, $\hat{G}(\omega,\mathbf{k})=\hat{G}^T(\omega,-\mathbf{k})$. Since the action Hessian represents the linear response operator and is the inverse of the Green's function \cite{lanczos1996linear}, \emph{i.e}., $\hat{H}(\omega,\mathbf{k})=\hat{G}^{-1}(\omega,\mathbf{k})$, the reciprocal symmetry of the Green's function directly implies the corresponding symmetry of the action Hessian,
\begin{align}\label{reciproc}
    \hat{H}(\omega,\mathbf{k})=\hat{H}^T(\omega,-\mathbf{k}).
\end{align}
Therefore, violation of this symmetry signals the breakdown of Lorentz reciprocity. Defining 
\begin{align}
    \hat{N}_k=\hat{K}^T\hat{C}_k-\hat{C}_k^T\hat{K},
\end{align}
since $\hat{N}_k(\omega,\mathbf{k})=-\hat{N}_k^T(\omega,-\mathbf{k})$, Eq.~\eqref{reciproc} holds only when $\hat{N}_k=0$. Noting that $\hat{K}$ is antisymmetric, namely $\hat{K}=-\hat{K}^T$, this requires 
\begin{align}\label{symCk}
    \hat{C}_k+\hat{C}_k^T=0,
\end{align}
namely $\hat{C}_k$ is antisymmetric. According to Eq.~\eqref{decompose}, this implies that the Lorentz reciprocity is broken when either effective isotropic chiral or pseudochiral coupling is emergent while preserved when only the omega-type coupling arises. 

\begin{figure}[t]
    \centering
    \includegraphics[width=\linewidth]{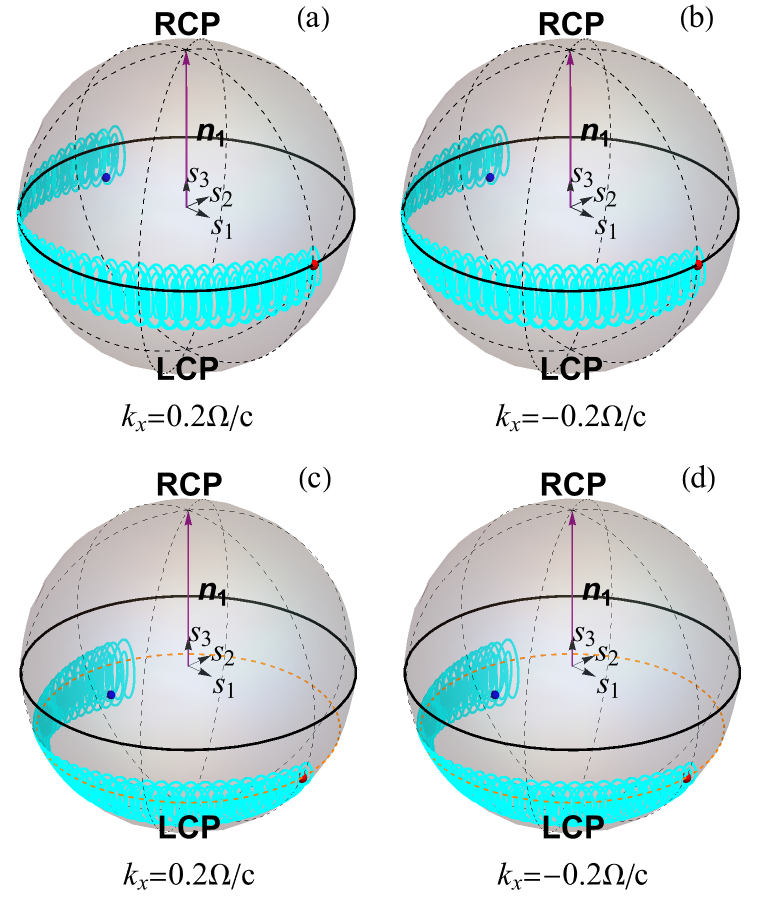}\caption{Trajectories of polarization states on the Poincaré sphere for propagation along the $x$-axis in the temporal metamaterial. The cyan curves represent the polarization trajectories, with the red and blue markers indicating the initial ($t=0$) and final ($t=200T$) states, respectively. The black solid circles indicate the equators, while the orange dashed circles denote the trajectories' latitudes, determined by the initial states. The purple arrows indicate the normalized Stokes vectors of the first effective eigenmodes. Panels (a) and (c) correspond to forward propagation, while panels (b) and (d) correspond to backward propagation. The temporal metamaterial parameters are the same as those used in Fig.~\ref{fig1}.}\label{fig2}
\end{figure}

\subsection{Unidirectional rotation of polarization state}
In anisotropic media, polarization evolution can be described by the continuous trajectory of the Stokes vector on the Poincaré sphere as a plane wave propagates \cite{kuratsuji1998maxwell,kuratsuji2021evolution,kravtsov2007stokes}. For our temporal metamaterials, within the effective-medium framework, let $\vec{\psi}_1$ and $\vec{\psi}_2$ denote the normalized Jones eigenvectors of the two effective eigenmodes. The polarization state at an arbitrary time can then be expressed as
\begin{align}
    \vec{\psi}(t)=c_1\vec{\psi}_1e^{-i\omega_1t}+c_2\vec{\psi}_2e^{-i\omega_2t},
\end{align}
where $\mathbf{k}\cdot\vec{\psi}_1=\mathbf{k}\cdot\vec{\psi}_2=0$ for transverse waves, $\vec{\psi}_m^\dag\vec{\psi}_n=\delta_{mn}$, $c_{1,2}$ are the expansion coefficients, and $\omega_{1,2}>0$ are the eigenfrequencies of the corresponding effective eigenmodes. The temporal evolution of the Jones vector is governed by
\begin{align}\label{pevolu}
i\partial_t\vec{\psi}=\hat{H}_{\rm pol}(\mathbf{k})\vec{\psi},
\end{align}
where $\hat{H}_{\rm pol}(\mathbf{k})$ is the Jones evolution generator \cite{kuratsuji1998maxwell}, given by
\begin{align}\label{generator}
\hat{H}_{\rm pol}(\mathbf{k})=\left(\begin{array}{cc}\vec{\psi}_1 & \vec{\psi}_2\end{array}\right)\left(\begin{array}{cc}\omega_1(\mathbf{k}) & 0 \\ 0 & \omega_2(\mathbf{k})\end{array}\right)\left(\begin{array}{c}\vec{\psi}_1^\dag \ \vec{\psi}_2^\dag\end{array}\right).
\end{align}
Unlike in conventional time-independent media, where polarization evolution is typically described as a function of the propagation coordinate $z$ \cite{kuratsuji1998maxwell}, polarization in our temporal metamaterials evolves in time at fixed wavevector $\mathbf{k}$.

According to the definition, the normalized three-component Stokes vector is given by \cite{kuratsuji1998maxwell}
\begin{align}
\mathbf{s}=\frac{\vec{\psi}^\dag\vec{\sigma}\vec{\psi}}{\vec{\psi}^\dag\vec{\psi}},
\end{align}
where $\vec{\sigma}=(\hat{\sigma}_z,\hat{\sigma}_x,\hat{\sigma}_y)$ is the reordered Pauli vector. With this convention, $s_3$ characterizes the polarization handedness, while $\phi=\frac{1}{2}\tan^{-1}(s_2/s_1)$ defines the orientation angle of the polarization ellipse in the transverse plane. The north and south poles of the Poincaré sphere correspond to right-circular polarization (RCP) and left-circular polarization (LCP), respectively, whereas points along the equator represent linear polarization states.

According to Eq.~\eqref{generator}, the Jones evolution generator can be decomposed as
\begin{align}
\hat{H}_{\rm pol}(\mathbf{k})=\overline{\omega}\hat{\sigma}_0+\frac{\Delta\omega}{2}\mathbf{n}_1(\mathbf{k})\cdot\vec{\sigma},
\end{align}
where $\overline{\omega}=(\omega_1+\omega_2)/2$ is the mean eigenfrequency, $\Delta\omega=\omega_2-\omega_1$ is the eigenfrequency splitting, and $\mathbf{n}_1(\mathbf{k})$ is the normalized Stokes vector of the first effective eigenmode, defined as
\begin{align}
\mathbf{n}_1(\mathbf{k})=\vec{\psi}_1^\dag\vec{\sigma}\vec{\psi}_1.
\end{align}
The temporal evolution of the Stokes vector then follows from the Jones equation as \cite{kuratsuji1998maxwell}
\begin{align}\label{dots}
\frac{d\mathbf{s}}{dt}
=\frac{i\vec{\psi}^\dag[\hat{H}_{\rm pol},\vec{\sigma}]\vec{\psi}}{\vec{\psi}^\dag\vec{\psi}}
=\Delta\omega\mathbf{n}_1(\mathbf{k})\times\mathbf{s}.
\end{align}
Equation~\eqref{dots} shows that the normalized Stokes vector undergoes uniform precession about the axis defined by $\mathbf{n}_1(\mathbf{k})$, with an angular rotation rate of magnitude $|\Delta\omega|$.

Since $\hat{H}(\omega,\mathbf{k})=\hat{H}(\omega,-\mathbf{k})$, Eq.~\eqref{linear1} directly implies that $\omega_{1,2}(\mathbf{k})=\omega_{1,2}(-\mathbf{k})$ and $\overline{\mathbf{A}}_{1,2}(\mathbf{k})=\overline{\mathbf{A}}_{1,2}(-\mathbf{k})$. Consequently, $\Delta\omega(\mathbf{k})=\Delta\omega(-\mathbf{k})$ and $\mathbf{n}_1(\mathbf{k})=\mathbf{n}_1(-\mathbf{k})$. It follows that the Stokes-vector rotation vector, $\Delta\omega\mathbf{n}_1$, remains unchanged under reversal of the propagation direction. Thus, the polarization evolution has the same rotational sense for $\mathbf{k}$ and $-\mathbf{k}$.

\begin{figure}[t]
    \centering
    \includegraphics[width=\linewidth]{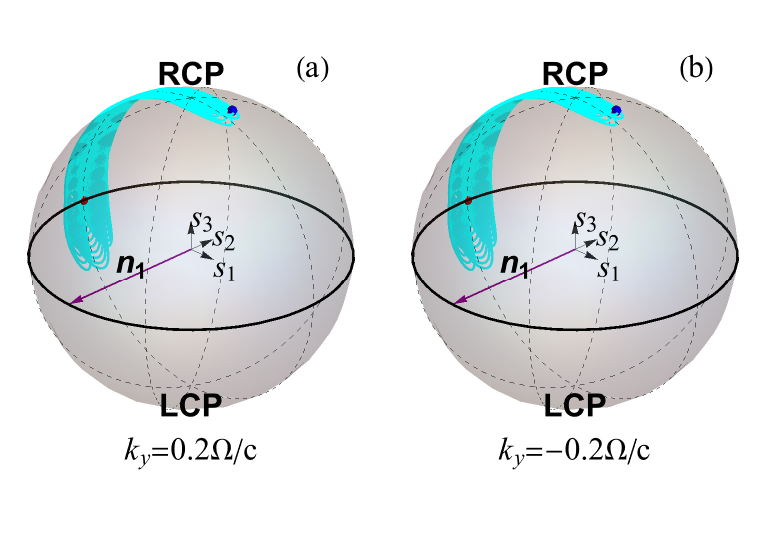}\caption{Trajectories of polarization states on the Poincaré sphere for propagation along the $y$-axis in the temporal metamaterial. The cyan curves represent the polarization trajectories, with the red and blue markers indicating the initial ($t=0$) and final ($t=200T$) states, respectively. The purple arrows indicate the normalized Stokes vectors of the first effective eigenmodes. Panels (a) and (b) correspond to forward and backward propagation, respectively. The temporal metamaterial parameters are the same as those used in Fig.~\ref{fig1}.}\label{fig3}
\end{figure}

Fig. \ref{fig2} shows the trajectories of the normalized Stokes vectors of the microscopic electromagnetic fields on the Poincaré sphere as a plane wave propagates along the \(x\)-axis in the temporal metamaterial. As established in the previous section, the temporal metamaterial exhibits an effective isotropic chiral coupling for propagation along the \(x\)-direction. The microscopic electromagnetic fields are calculated rigorously by iteratively solving Maxwell’s equations in the time domain. In an isotropic chiral medium, the eigenmodes are RCP and LCP waves; accordingly, \(\mathbf{n}_1\) points toward the north pole of the Poincaré sphere. When the incident plane wave is initially linearly polarized, its normalized Stokes vector starts on the equator. As shown in Figs.~\ref{fig2}(a) and \ref{fig2}(b), for propagation in either direction, the normalized Stokes vector rotates clockwise around \(\mathbf{n}_1\), while remaining close to the equator. This behavior indicates that the wave largely preserves its linear polarization while its polarization angle evolves continuously. We refer to this phenomenon as the temporal Faraday effect, as introduced in our previous work \cite{wang2026temporal}.

For an initially elliptically polarized wave, as shown in Figs. \ref{fig2}(c) and \ref{fig2}(d), the normalized Stokes vector continues to rotate clockwise around \(\mathbf{n}_1\), for both forward and backward propagation, while remaining close to a fixed meridian. Moreover, over the same evolution time, the angular displacement of the normalized Stokes vector is identical to that in Figs. \ref{fig2}(a) and \ref{fig2}(b), demonstrating that the initial polarization state does not affect the rotation rate, \(\Delta\omega\).

\begin{figure}[t]
    \centering
    \includegraphics[width=\linewidth]{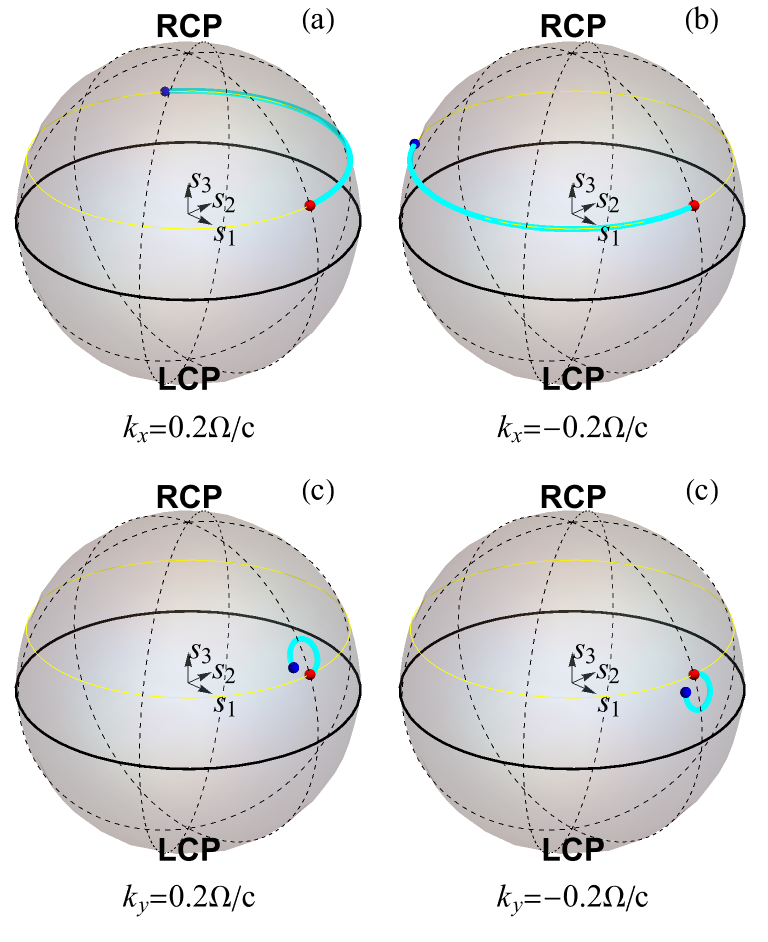}\caption{Trajectories of polarization states on the Poincaré sphere in static, $\mathbf{k}$-independent isotropic chiral and pseudochiral media. The cyan curves represent the polarization trajectories, with the red and blue markers indicating the initial and final states, respectively. (a) Isotropic chiral medium with $\hat{\varepsilon}=\varepsilon_0\hat{I}_3$, $\hat{\nu}=\mu_0^{-1}\hat{I}_3$, and $\hat{C}\perp^{(x)}=0.1\hat{\sigma}_0$. (b) Isotropic pseudochiral medium with $\hat{\varepsilon}=\varepsilon_0\hat{I}_3$, $\hat{\nu}=\mu_0^{-1}\hat{I}_3$, and $\hat{C}\perp^{(y)}=0.1\hat{\sigma}_z$.}\label{fig4}
\end{figure}

The trajectories of the normalized Stokes vectors of the microscopic electromagnetic fields on the Poincaré sphere for a plane wave propagating along the \(y\)-axis in the temporal metamaterial are shown in Fig. \ref{fig3}. Along the \(y\)-direction, the temporal metamaterial exhibits an anisotropic pseudochiral coupling, with the anisotropy arising from higher-order terms neglected in \(\mathcal{L}_{\rm eff}\). As shown in Fig. \ref{fig3}, the normalized Stokes vector rotates clockwise around \(\mathbf{n}_1\) for both forward and backward propagation. Notably, because the pseudochiral coupling strength satisfies \(\gamma^2 k_yc/(2\Omega)\ll 1\), \(\mathbf{n}_1\) lies close to the equatorial plane. Furthermore, the anisotropy induced by the higher-order terms causes \(\mathbf{n}_1\) to lie nearly parallel to the \(s_2\)-axis, corresponding to a \(45^\circ\) rotation of the principal axes of the effective medium relative to the \(xz\)-plane. A rigorous calculation of \(\mathbf{n}_1\) is using the Floquet matrix method \cite{wang2018photonic,lu2025generalized}, where the Floquet states are calculated accurately, and the corresponding Jones eigenvectors of the effective medium are obtained by temporal averaging. Note that the Stokes-vector rotation here is primarily driven by anisotropy rather than the weak pseudochirality. Achieving isotropic pseudochirality is nontrivial, and the high-frequency oscillations of the microscopic fields, known as micromotion, can further obscure the polarization rotation induced by the weak pseudochiral response.

This direction-independent polarization rotation is a hallmark of nonreciprocal behavior. Because the polarization rotates in the same sense for forward and backward propagation, a wave that traverses the medium in one direction and then returns along the reverse path accumulates additional polarization rotation rather than retracing its previous evolution. Consequently, the polarization state generally does not return to its initial state after a round trip. This behavior contrasts with reciprocal optical activity, where the polarization evolution reverses upon reversing the propagation direction, allowing the accumulated rotation to cancel in a round trip. Moreover, isotropic omega-type coupling does not induce any evolution of the normalized Stokes vector and therefore does not give rise to nonreciprocity.

When the bianisotropic coupling tensor \(\hat{C}_{\mathbf{k}}\) is an even function of \(\mathbf{k}\), as in conventional static bianisotropic media where \(\hat{C}_{\mathbf{k}}\) is independent of \(\mathbf{k}\), Eq.~\eqref{linear2} yields \(\hat{H}(\omega,\mathbf{k})=\hat{H}^*(\omega,-\mathbf{k})\neq\hat{H}(\omega,-\mathbf{k})\). Meanwhile, the even parity of \(\hat{C}_{\mathbf{k}}\) ensures that \(\hat{H}(\omega,\mathbf{k})=\hat{H}^{T}(\omega,-\mathbf{k})\), thereby satisfying Lorentz reciprocity. In this case, reversing the wave vector also reverses the sign of \(\Delta\omega\), causing the normalized Stokes vectors of forward- and backward-propagating waves to rotate in opposite directions.

In Fig. 4, we examine the trajectories of the normalized Stokes vectors of plane waves propagating through conventional static isotropic chiral and pseudochiral media, namely the bianisotropic coupling tensors are $\mathbf{k}$-independent. As shown, for these reciprocal bianisotropic media, the normalized Stokes vectors indeed rotate in opposite directions for forward and backward propagation.

\section{Conclusions}
In conclusion, we have developed an action-based framework for describing nonlocal bianisotropy and Lorentz reciprocity in temporally modulated anisotropic media. By performing a multiscale expansion of the microscopic Lagrangian density, we derived an effective electromagnetic action in which the first-order spatial correction gives rise to a wavevector-odd bianisotropic coupling tensor. This unified description encompasses chiral, pseudochiral, and omega-type responses, despite the underlying symmetric electric and magnetic response tensors. By analyzing the Hessian of the effective action, we established a systematic approach to identifying the reciprocity properties of these effective couplings. Furthermore, our analysis of the normalized Stokes vectors on the Poincaré sphere reveals distinct polarization-evolution behaviors associated with different coupling symmetries, providing insight into the relationship between directional polarization rotation and nonreciprocity. In particular, the direction-independent sense of polarization rotation associated with Floquet-induced chirality provides a distinctive signature of nonreciprocal polarization evolution, whereas conventional reciprocal optical activity exhibits opposite rotation senses for opposite propagation directions. These findings establish a unified theoretical foundation for understanding effective bianisotropy, polarization dynamics, and reciprocity in temporal metamaterials, and may guide the design of dynamically reconfigurable photonic systems with tailored polarization and nonreciprocal responses.

\begin{acknowledgments}
This work was supported by the National Natural Science Foundation of China (NSFC) (No. 12674470 and No. 12434017) and the Key Project of the National Key R$\&$D program of China (No. 2022YFA1404500).
\end{acknowledgments}

\nocite{*}

\bibliography{apssamp}

\end{document}